\documentclass{article}[10pt]
\usepackage{graphicx,subfigure}
\usepackage{epsfig}
\usepackage{algorithm}
\usepackage[noend]{algorithmic}
\usepackage{setspace}
\usepackage{amsthm}
\usepackage{amsfonts}
\usepackage{amsmath}
\usepackage{color}
\usepackage{array}
\usepackage{setspace}
\usepackage{url}

\usepackage{xr}

\begin{document}

\title{Comparative Assembly Hubs: Web Accessible Browsers for Comparative Genomics}



\maketitle


Ngan Nguyen$^{1*}$, Glenn Hickey$^{1*}$, Brian J. Raney$^{1*}$, Joel Armstrong$^1$, Hiram Clawson$^1$,  Ann Zweig$^1$,  Jim Kent$^1$, David Haussler$^{1,2}$, Benedict Paten$^{1**}$

$^1$ Center for Biomolecular Sciences and Engineering, CBSE/ITI, UC Santa Cruz, 1156 High St, Santa Cruz, CA 95064, USA.\\
$^2$ Howard Hughes Medical Institute, University of California, Santa Cruz, CA 95064, USA.\\

* These authors contributed equally to this work.
** Corresponding author: benedict@soe.ucsc.edu


\begin{abstract}
We introduce a pipeline to easily generate collections of web accessible UCSC genome browsers interrelated by an alignment. Using the alignment, all annotations and the alignment itself can be efficiently viewed with reference to any genome in the collection, symmetrically. A new, intelligently scaled alignment display makes it simple to view all changes between the genomes at all levels of resolution, from substitutions to complex structural rearrangements, including duplications. 
\end{abstract}

Visualization is key to understanding functional and comparative genomic information.  Genome browsers are therefore critical to the study of biology, providing accessible resources for displaying annotations and alignments.
The UCSC Genome Browser \cite{Meyer:2013hd} is one of the most popular, but creating a browser within it previously required significant resources, as it was necessary to create a mirror site to host the browser separately, or to work with the staff of the genome browser to create a browser within their main site. 
Assembly hubs, which build on the successful track hub model \cite{trackhub}, make it easy to generate a UCSC browser simply by hosting the data in the form of flat-files on any publicly addressable URL \cite{Meyer:2013hd}. This avoids users having to install and configure the substantial browser code base on their machines and, by using user hosting, makes updating straightforward.
This work extends assembly hubs to allow users to quickly create ``comparative assembly hubs'', a framework of genome browsers and annotations interrelated by an alignment.
Integral to this framework, we have added snake tracks, to allow all structural variations to be viewed. Snake tracks are analogous to a linearization of the popular Circos plots \cite{krzywinski2009circos, Nielsen:2010ea}, but implement intelligent alignment scaling, making them useful at all levels of detail.


Our software pipeline is comprised of three components, for which we provide an overall distribution that can be downloaded and installed with just three commands (see Methods), and which should work on unix distributions such as Linux, BSD and OS X. The first component is the Cactus alignment program \cite{Paten:2011fv}, which takes as input a set of genome sequences and outputs a genome multiple sequence alignment in Hierarchical Alignment (HAL) format \cite{Hickey:2013jo}. The second component is HAL tools, which provides a series of command line tools and C/C++ APIs for manipulating HAL files and building comparative assembly hubs. The final component is the snake track display, which is part of the UCSC Genome Browser code base \cite{Meyer:2013hd}, and which provides visualization of alignments directly from HAL files. 


The pipeline is run in two stages (see Methods). Firstly, Cactus is run to generate the HAL alignment file, then the hal2AssemblyHub script (in the HAL tools package) builds the comparative assembly hub using the HAL file and any set of annotation files provided, either in BED or WIG format (\url{http://genome.ucsc.edu/FAQ/FAQformat. html}). This script takes care of converting the base annotation files into the display scaleable bigBed and bigWig formats, and translates these annotations (optionally) via a process of alignment lift-over \cite{Zhu:2007jp}, to all the other genomes. At the end of this process a directory is created that contains the necessary files, using compressed formats for minimal space usage. The location of the `hub.txt' file, addressable as a public URL, is pasted into the UCSC browser hub page to view the browsers. 



\begin{figure} 
\centering
\includegraphics[width=450pt]{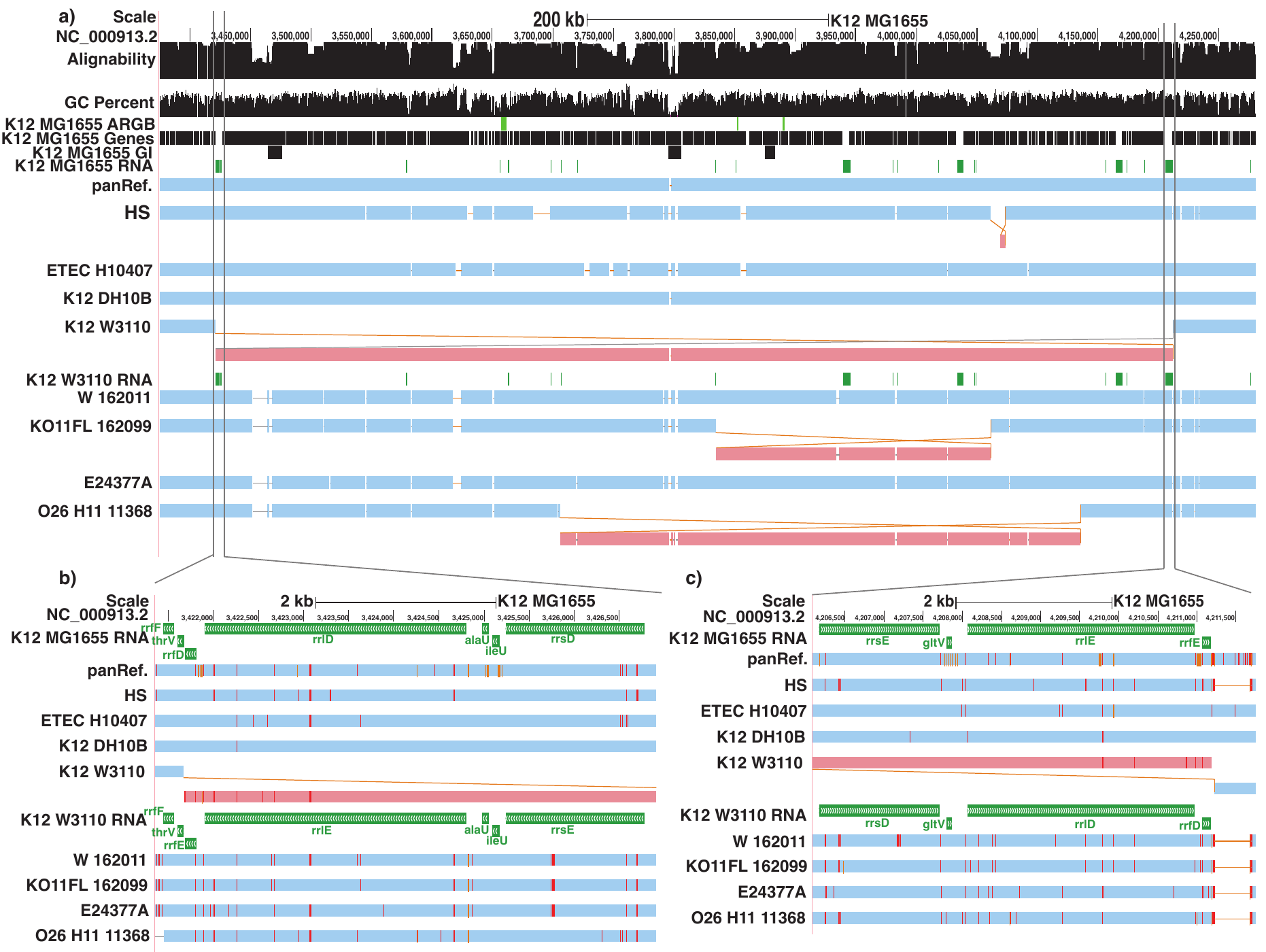}
\caption{An example \textit{E. coli} comparative assembly hub with \textit{E. coli} K12 MG1655 as the reference browser. 
 The top browser screenshot (\textbf{a}) shows a ~900kb region with a known large inversion (light red) in the closely related strain K12 W3110, which is flanked by homologous (with opposite orientations) ribosomal RNA operons \textit{rrnD} and \textit{rrnE} \cite{Hill:1981tj, Hayashi:2006dj}, and is the result of recombination between them. (\textbf{b-c}) Zoom-in of the K12 W3110 inversion left and right boundaries, respectively, showing operon \textit{rrnE} of K12 W3110 (`K12\_W3110 RNA' track, in green, which is K12 W3110 ncRNA annotation track lifted-over to K12 MG1655) aligned to operon \textit{rrnD} of K12 MG1655 (`K12\_MG1655 RNA' track, also in green) on the left and operon \textit{rrnD} of K12 W3110 aligned to operon \textit{rrnE} of K12 MG1655 on the right. Zoomed in, SNPs and query insertions are visible. The text on the screenshots was adjusted for better readability.
}
\label{fig:k12inversion}
\end{figure}

The pipeline builds one browser for each input genome, and, in addition, any ancestral or pan- genomes that were imputed by Cactus during the alignment process (Paten \textit{et al.}, submitted). 
To illustrate by example, Figure \ref{fig:k12inversion} shows a comparative assembly hub containing \textit{E. coli} and \textit{Shigella} genomes (see Methods), displayed upon one of the \textit{E. coli} reference genomes, K12 MG1655, with snake tracks, BED and WIG annotations and lifted-over BED annotations. The top most tracks are K12 MG1655 annotations, including Alignability (number of genomes mapped to each position), GC\%, Antibiotic Resistance Database (ARDB), Genes, Genomic Islands (GI) and non-coding RNAs (rRNA and tRNA). Below these are snake tracks, showing the alignment of the genome to a subset of the other genomes and lifted-over ncRNA annotation track (track K12 W3110 RNA) of \textit{E. coli} K12 W3110.

Each snake track shows the relationship between the chosen browser genome, termed the reference (genome), and another genome, termed the query (genome). The snake display is capable of showing all possible types of structural rearrangement (see Methods). 
In \textit{full} display mode (snake tracks in Figure \ref{fig:k12inversion}), it can be decomposed into two primitive drawing elements, segments, which are the colored rectangles, and adjacencies, which are the lines connecting the segments. Segments represent subsequences of the query genome aligned to the given portion of the reference genome. Adjacencies represent the covalent bonds between the aligned subsequences of the query genome.  Segments can be configured to be colored by chromosome, strand (as shown) or left a single color.  
Red tick-marks within segments represent substitutions with respect to the reference, shown in windows of the reference of (by default, user configurable) up to 50 kilo-bases (Figure \ref{fig:k12inversion}(b-c)). Zoomed in to the base-level these substitutions are labeled with the non-reference base. An insertion in the reference relative to the query creates a gap between abutting segment sides that is connected by an adjacency. An insertion in the query relative to the reference is represented by an orange tick mark that splits a segment at the location the extra bases would be inserted, or by coloring an adjacency orange, indicating that there are unaligned bases between the two segment ends it connects. More complex structural rearrangements create adjacencies that connect the sides of non-abutting segments in a natural fashion, for example, Figure \ref{fig:k12inversion} shows the known large inversion in K12 MG1655's closely related strain K12 W3110 \cite{Hill:1981tj, Hayashi:2006dj}.



\begin{figure}
\centering
\includegraphics[width=450pt]{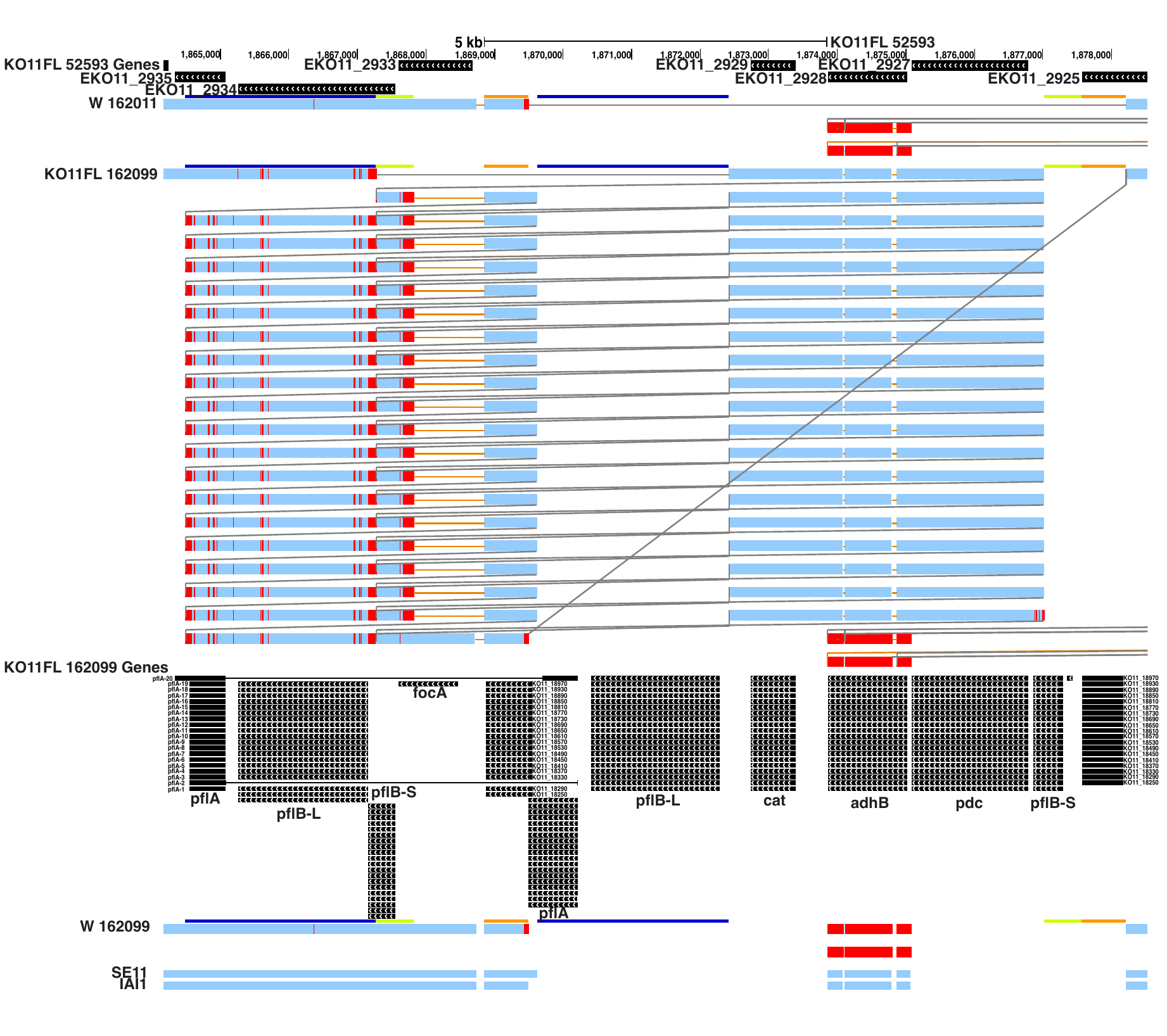}
\caption{A browser screenshot showing the \textit{pdc-adhB-cat} tandem repeat region of \textit{E. coli} KO11FL 162099 \cite{Turner:2012dv} displayed along the genome of \textit{E. coli} KO11FL 52593. 
The colored  horizontal bars on top of each snake track indicate duplications in KO11FL 52593 (two copies of each gene \textit{pflA} (dark blue), \textit{pflB-L} (dark blue), \textit{pflB-S} (light green) and KO11\_18XXX (orange)). There is a large deletion in the parent strain W 162011 as this strain does not contain the \textit{pdc-adhB-cat} insert. Following the snake track of KO11FL 162099, there are 20 copies of (\textit{pflA}, \textit{pflB-L}, \textit{cat}, \textit{adhB}, \textit{pdc}, \textit{pdfB-S} and \textit{KO11-18XXX}). Note that since KO11FL 52593 has two copies of \textit{pflA}, \textit{pflB-L}, \textit{pflB-S} and KO11\_18XXX, the display arbitrarily picks one copy of each to map corresponding KO11FL 162099 orthologous genes to. The text on the screenshot was adjusted for better readability.}
\label{fig:dups}
\end{figure}

Duplications within the query genome create extra segments that overlap along the reference genome axis. For example, Figure \ref{fig:dups} shows a tandem repeat region of \textit{E. coli} KO11FL 162099 displayed along the genome of \textit{E. coli} KO11FL 52593 (52593 was engineered by chromosomal insertion of the \textit{Zymomonas mobilis} \textit{pdc}, \textit{adhB} and \textit{cat} genes into \textit{E. coli} W for ethanol production purpose \cite{Ohta:1991tp}; 162099 is a derivative of 52593 and contains 20 tandem copies of the inserted \textit{pdc-adhB-cat} genes\cite{Turner:2012dv}). To show regions where the query segments align to multiple locations within the reference, for each snake track we draw colored coded sets of lines along the reference genome axis that indicate self homologies (intervals of the reference genome that align to other intervals of the reference genome), and, to maintain the semantics of the snake, align any query segments that align to these regions arbitrarily to just one copy of the reference (Figure \ref{fig:dups}). 
%


The \textit{pack} display option can be used to display a snake track in more limited vertical space. It eliminates the adjacencies from the display and forces the segments onto as few rows as possible, given the constraint of still showing duplications in the query sequence (e.g. track W 162099 of Figure \ref{fig:dups}). The \textit{dense} display further eliminates these duplications so that a snake track is compactly represented along just one row (e.g. tracks SE11 and IAI1 of Figure \ref{fig:dups}). 
Clicking on segments moves to the corresponding region in the query genome, making it simple to navigate between references. Various mouse-overs are implemented to show the sizes of display elements, and the snakes and annotations can be reordered by dragging. A ``hub central'' configuration page makes managing the large number of possible snake and lifted-over annotations easy. Export of subregions of the alignment and track intersections can be made via the UCSC table browser \cite{Karolchik:2004bc}. In the methods we describe a novel algorithm that, analogously to the bigBed and bigWig formats, creates the scaleable levels of alignment detail, and show that this achieves near constant time page reloads at all possible zoom levels.  We have tested comparative assembly hubs with clades of mammalian genomes, demonstrating that they are practical even for large animal genomes. 



\section*{Acknowledgements}

We would like to thank the Howard Hughes Medical Institute, Dr. and Mrs. Gordon Ringold, NIH grant 2U41 HG002371-13 and NHGRI/NIH grant 5U01HG004695 for providing funding.

\bibliographystyle{naturemag}
\bibliography{ecoli}

\begin{thebibliography}{10}
\expandafter\ifx\csname url\endcsname\relax
  \def\url#1{\texttt{#1}}\fi
\expandafter\ifx\csname urlprefix\endcsname\relax\def\urlprefix{URL }\fi
\providecommand{\bibinfo}[2]{#2}
\providecommand{\eprint}[2][]{\url{#2}}

\bibitem{Meyer:2013hd}
\bibinfo{author}{Meyer, L. R.~L.} \emph{et~al.}
\newblock \bibinfo{title}{{The UCSC Genome Browser database: extensions and
  updates 2013.}}
\newblock \emph{\bibinfo{journal}{Nucleic Acids Research}}
  \textbf{\bibinfo{volume}{41}}, \bibinfo{pages}{D64--D69}
  (\bibinfo{year}{2013}).

\bibitem{trackhub}
\bibinfo{author}{Raney, B.} \emph{et~al.}
\newblock \bibinfo{title}{{Track Data Hubs enable visualization of user-defined
  genome-wide annotations on the UCSC Genome Browser}}.
\newblock \emph{\bibinfo{journal}{Bioinformatics}}  (\bibinfo{year}{2013}).
\newblock \bibinfo{note}{In-press}.

\bibitem{krzywinski2009circos}
\bibinfo{author}{Krzywinski, M.} \emph{et~al.}
\newblock \bibinfo{title}{Circos: an information aesthetic for comparative
  genomics}.
\newblock \emph{\bibinfo{journal}{Genome research}}
  \textbf{\bibinfo{volume}{19}}, \bibinfo{pages}{1639--1645}
  (\bibinfo{year}{2009}).

\bibitem{Nielsen:2010ea}
\bibinfo{author}{Nielsen, C.~B.}, \bibinfo{author}{Cantor, M.},
  \bibinfo{author}{Dubchak, I.}, \bibinfo{author}{Gordon, D.} \&
  \bibinfo{author}{Wang, T.}
\newblock \bibinfo{title}{{Visualizing genomes: techniques and challenges.}}
\newblock \emph{\bibinfo{journal}{Nature methods}}
  \textbf{\bibinfo{volume}{7}}, \bibinfo{pages}{S5--S15}
  (\bibinfo{year}{2010}).

\bibitem{Paten:2011fv}
\bibinfo{author}{Paten, B.} \emph{et~al.}
\newblock \bibinfo{title}{{Cactus: Algorithms for genome multiple sequence
  alignment.}}
\newblock \emph{\bibinfo{journal}{Genome research}}
  \textbf{\bibinfo{volume}{21}}, \bibinfo{pages}{1512--1528}
  (\bibinfo{year}{2011}).

\bibitem{Hickey:2013jo}
\bibinfo{author}{Hickey, G.}, \bibinfo{author}{Paten, B.},
  \bibinfo{author}{Earl, D.}, \bibinfo{author}{Zerbino, D.} \&
  \bibinfo{author}{Haussler, D.}
\newblock \bibinfo{title}{{HAL: a hierarchical format for storing and analyzing
  multiple genome alignments.}}
\newblock \emph{\bibinfo{journal}{Bioinformatics}}
  \textbf{\bibinfo{volume}{29}}, \bibinfo{pages}{1341--1342}
  (\bibinfo{year}{2013}).

\bibitem{Zhu:2007jp}
\bibinfo{author}{Zhu, J.} \emph{et~al.}
\newblock \bibinfo{title}{{Comparative genomics search for losses of
  long-established genes on the human lineage.}}
\newblock \emph{\bibinfo{journal}{PLoS Computational Biology}}
  \textbf{\bibinfo{volume}{3}}, \bibinfo{pages}{e247--e247}
  (\bibinfo{year}{2007}).

\bibitem{Hill:1981tj}
\bibinfo{author}{Hill, C.~W.} \& \bibinfo{author}{Harnish, B.~W.}
\newblock \bibinfo{title}{{Inversions between ribosomal RNA genes of
  \textit{Escherichia coli}.}}
\newblock \emph{\bibinfo{journal}{Proceedings of the National Academy of
  Sciences of the United States of America}} \textbf{\bibinfo{volume}{78}},
  \bibinfo{pages}{7069--7072} (\bibinfo{year}{1981}).

\bibitem{Hayashi:2006dj}
\bibinfo{author}{Hayashi, K.~K.} \emph{et~al.}
\newblock \bibinfo{title}{{Highly accurate genome sequences of
  \textit{Escherichia coli} K-12 strains MG1655 and W3110.}}
\newblock \emph{\bibinfo{journal}{Molecular Systems Biology}}
  \textbf{\bibinfo{volume}{2}}, \bibinfo{pages}{2006--0007}
  (\bibinfo{year}{2006}).

\bibitem{Ohta:1991tp}
\bibinfo{author}{Ohta, K.~K.}, \bibinfo{author}{Beall, D. S.~D.},
  \bibinfo{author}{Mejia, J. P.~J.}, \bibinfo{author}{Shanmugam, K. T.~K.} \&
  \bibinfo{author}{Ingram, L. O.~L.}
\newblock \bibinfo{title}{{Genetic improvement of \textit{Escherichia coli} for
  ethanol production: chromosomal integration of \textit{Zymomonas mobilis}
  genes encoding pyruvate decarboxylase and alcohol dehydrogenase II.}}
\newblock \emph{\bibinfo{journal}{Applied and Environmental Microbiology}}
  \textbf{\bibinfo{volume}{57}}, \bibinfo{pages}{893--900}
  (\bibinfo{year}{1991}).

\bibitem{Turner:2012dv}
\bibinfo{author}{Turner, P. C.~P.} \emph{et~al.}
\newblock \bibinfo{title}{{Optical mapping and sequencing of the
  \textit{Escherichia coli} KO11 genome reveal extensive chromosomal
  rearrangements, and multiple tandem copies of the \textit{Zymomonas mobilis}
  \textit{pdc} and \textit{adhB} genes.}}
\newblock \emph{\bibinfo{journal}{Journal of Industrial Microbiology {\&}
  Biotechnology}} \textbf{\bibinfo{volume}{39}}, \bibinfo{pages}{629--639}
  (\bibinfo{year}{2012}).

\bibitem{Karolchik:2004bc}
\bibinfo{author}{Karolchik, D.} \emph{et~al.}
\newblock \bibinfo{title}{{The UCSC Table Browser data retrieval tool.}}
\newblock \emph{\bibinfo{journal}{Nucleic Acids Research}}
  \textbf{\bibinfo{volume}{32}}, \bibinfo{pages}{D493--D496}
  (\bibinfo{year}{2004}).

\bibitem{repeatMasker}
\bibinfo{author}{Smit, A. F.~A.}, \bibinfo{author}{Hubley, R.} \&
  \bibinfo{author}{Green, P.}
\newblock \bibinfo{title}{{RepeatMasker.} open-3.0}
  (\bibinfo{year}{1996-2013}).
\newblock \urlprefix\url{http://www.repeatmasker.org}.

\end{thebibliography}

\newpage

\section*{Methods} 

\subsection*{Availability and Installation}

The source code is all open-source and available from: \url{https://github.com/glennhickey/progressiveCactus}, which also gives the simple installation instructions and support contact information, should users experience difficulty. The browser sessions of Figures \ref{fig:k12inversion} and \ref{fig:dups} are available at \url{http://tinyurl.com/lu6o4sj} and \url{http://tinyurl.com/jwnp3p5}, respectively.

\subsection*{Alignment Assembly Hub Pipeline}
From a set of input genome sequences (and any available annotations), users can create the comparative assembly hub using the two following commands:

1. \textit{runProgressiveCactus.sh $<$seqFile$>$ $<$workDir$>$ $<$outputHalFile$>$}\\
2. \textit{hal2assemblyHub.py $<$halFile$>$ $<$outDir$>$ --bedDirs $<$annotationDirs$>$ --lod}

Command (1) generates the multiple sequence alignment, which is stored in the HAL format to the specified output file $outputHalFile$. $seqFile$ contains Newick-formatted phylogenetic tree of the input genomes (optional) and paths to the sequence FASTA files. $workDir$ specifies the working directory. More details can be found in Progressive Cactus Manual (\url{https://github.com/glennhickey/progressiveCactus}).

Command (2) produces necessary data and files for creating the comparative assembly hubs through the UCSC genome browser. $halFile$ is the HAL-formatted MSA file, which is the output ($outputHalFile$) from command (1). $outDir$ is the output directory where all the generated files are written into. Among the output files is a file named ``hub.txt'', which the users will upload to the UCSC genome browser (similarly to how a track hub is created \cite{trackhub}, see \url{http://genome.ucsc.edu/goldenPath/help/hgTrackHubHelp.html} for more details) and the comparative assembly hubs will be created. $annotationDirs$ is a comma separated list of directories containning the annotation files, one directory per annotation type (e.g genes, pathogenic regions, antiobiotic resistance regions). Option \textit{--lod} is specified to compute the levels of detail, which is recommended for large datasets. For parallelism and job management, $hal2assemblyHub.py$ uses jobTree (\url{https://github.com/benedictpaten/jobTree}), which is installed as part of Progressive Cactus installation process. Users can specify different jobTree options to speed up the running time.

In this work, we generated two different \textit{E. coli}/\textit{Shigella} comparative assembly hubs.  One with duplications allowed (\url{http://compbio.soe.ucsc.edu/reconstruction/ecoliComparativeHubs/ecoliWithDups/hub/hub.txt}) and one with duplications disallowed \\ (\url{http://compbio.soe.ucsc.edu/reconstruction/ecoliComparativeHubs/ecoliNoDups/hub/hub.txt}).

Each of the two hubs were generated by the two following commands:

1. \textit{runProgressiveCactus.sh --legacy --configFile config.xml --maxThreads 24 --ktType snapshot seqFile.txt outdir outdir/alignment.hal}\\
2. \textit{hal2assemblyHub.py alignment.hal outHubDir --maxThreads 24 --lod --bedDirs Genes,RNA,GI,PI,PathogenicGenes,ARGB --rmskDir rmskTracks --gcContent --alignability --conservation conservationRegions.bed --conservationGenomeName reference --conservationTree tree.nw --tree tree.nw --rename shortnames.txt --hub ecoliCompHub --shortLabel EcoliCompHub --longLabel ``Escherichia coli Comparative Assembly Hub''}

All related files can be found at \url{http://compbio.soe.ucsc.edu/reconstruction/ecoliComparativeHubs}, under directories ``ecoliWithDups'' and ``ecoliNoDups'', respectively. For more details of the options, please see the $hal2assemblyHub$ documentation at \url{https://github.com/glennhickey/hal}. 

\subsection*{Genome Sequence and Annotation Data}
\label{sec:data}
Nucleotide sequences of 57 \textit{E. coli} and 9 \textit{Shigella} spp. complete genomes were downloaded from the NCBI ftp site (\url{ftp://ftp.ncbi.nlm.nih.gov/genomes/Bacteria/all.fna.tar.gz}, January 2013). The sequences were repeat-masked using RepeatMasker \cite{repeatMasker} with the `\emph{-xsmall}' option and otherwise default settings. The repeat-masked sequences were used as inputs to construct the MSA. Other outputs of RepeatMasker were converted into bigBed  format to build the ``Repetitive Elements'' track for each genome (\url{http://genomewiki.ucsc.edu/index.php/RepeatMasker}). For the 9 genomes ATCC 873, DH1 161951, KO11FL 162099, KO11FL 52593, O104 H4 2009EL 2050, O104 H4 2009EL 2071, O104 H4 2011C 3493, UM146, BL21 Gold DE3 pLysS AG, we used the reverse complement of their assemblies as the majority portion of those assemblies aligned to the reverse strand of other (57) genomes.

Gene, protein and non-coding RNA annotations for each genome were also obtained from NCBI (\url{ftp://ftp.ncbi.nlm.nih.gov/genomes/Bacteria/all.gff.tar.gz}, \url{ftp://ftp.ncbi.nlm.nih.gov/genomes/Bacteria/all.faa.tar.gz} and \url{ftp://ftp.ncbi.nlm.nih.gov/genomes/Bacteria/all.rnt.tar.gz}, respectively). 

\subsection*{Snake Display Algorithm}

A snake is a way of viewing a set of pairwise gap-less alignments that may
overlap on both the reference and query genomes.  Alignments are always represented as being on
the positive strand of the reference species, but can be on either
strand on the query sequence.

A snake plot puts all the query segments within a reference chromosome range
on a set of one or more levels.   All the segments on a level are on the
same strand, do not overlap in reference coordinate space, and are in the
same order and orientation in both sequences.   This is the same
requirement as the alignments in a chain on the UCSC browser.   Before the
algorithm is started,  all the segments are sorted by their starting
coordinate on the query, and the current level is set to one.    Then
in a recursive fashion,  the algorithm places the first segment on the
current list on the current level, and then adds all the rest of the
segments on the list that will fit onto the current level with the
requirements that all the segments on a level are on the same strand,
and that the proposed segment be non-overlapping and have a reference
start address that's greater than the reference end address of the
previously added segment on that level.   All segments that won't fit on
the current level are then added to subsequent levels following the
same rules.
Once all the segments have been assigned a level, lines are drawn
between the segments to show the adjacencies in the list when sorted by
query start address.

\subsection*{Alignment Graph Levels of Detail}

\subsubsection*{Interpolating Alignment Graphs}

A HAL graph  organizes the genomes it contains phylogenetically.  The pairwise alignment between any pair of genomes in the graph can be inferred by following the path between them \cite{Hickey:2013jo}.  The result is a set of gapless aligned blocks between the two genomes.  Algorithmically, such queries are simple graph traversals.  Practically, they can be quite difficult to compute in real time, as required for visualization, due to the data size involved.  For instance, a mammalian chromosome will typically be decomposed into up to 100 million segments in a HAL graph.  Computing the alignment between such a chromosome with a species that is 10 nodes away on the tree could require roughly 10 billion segments to be visited.  This is more than could be processed in real time, even if they were all stored in the cache.   Such a query would also produce far more information than could ever be usefully displayed on a screen.  We develop a system for pre-generating interpolated HAL graphs that will store only as much information as visible on the screen at different levels of detail.  Pseudocode for the interpolation procedure is provided in Algorithm \ref{alg:interpolate}, and each step is described in more detail below. 

\subsubsection*{Problem Definition}

We begin by briefly introducing the definition of a sequence graph (more details in \cite{Paten:2011fv}).  In a sequence graph $G$, each sequence $s \in S$ in the alignment (chromosome, contig, etc) is represented by a string of DNA, which is in turn partitioned into segments.  All homologous segments are grouped together into maximal gapless alignment blocks. Each segment within an alignment block is associated with a strand identifier to specify whether the forward or reverse strand of the segment is being aligned. Let $|G|$ refer to the number of blocks in $G$.  When comparing two different sequence graphs of the same input data, $G_1$ and $G_2$, we define $\Delta(G_1, G_2)$ to be the sum of pairwise base homologies induced by the blocks of $G_1$ and not $G_2$ with those induced by $G_2$ and not $G_1$. 

Sequence graphs are conceptually equivalent to HAL graphs. Using this equivalence, we define the interpolation problem as follows.

Given an a sequence graph $G$ and bound $K$, compute a sequence graph $G'$ such that $|G'| \leq K$ and $\Delta(G',G)$ is minimal.  In practice, $K$ is a funciton of the number of pixels in the browser display and is 100 by default.

Due to the size of the search space (all graphs with $\leq K$ blocks), we use the following sampling based solution.

\subsubsection*{Sampling The Column Graph}

We use a down-sampling algorithm based on the simplifying assumption that alignment block lengths are roughly uniform: if the total length of all blocks in $G$ is $L_{tot}$, then we expect that each block in $G'$ will have length approximately $L_{block} = L_{tot}/K$.  The first step of the algorithm is to therefore to sample an initial graph $G_0$ from $G$ by sampling every $L_{block}$ bases of each $s \in S$ and extracting the block of length one from $G$ into $G_0$ (if it hasn't been added already).  

In order to keep the number of sampled blocks proportional to $K$, we disregard sampled blocks whose maximum distance (along any segment) to any block already sampled is less than $L_{block}$. 

\subsubsection*{Extending The Column Graph}

Unless $L_{block}=1$, $G_0$ will not necessarily be a valid sequence graph since it will not contain all bases in $S$.  We therefore greedily extend each block in $G_0$ using the following rules, creating $G_1$.  We define a segment $e$ as a closed interval $(i,j)$ where $i \leq j$ on sequence $s \in S$. If $e$ is on the forward strand in its containing block and $s[j+1]$ exists and is not already present in $G_1$, or if $e$ is on the reverse strand and $s[i-1]$ exists and is not already in $G_1$, then $e$ can be extended to the right.  A similar check can determine if $e$ can be extended left.  $G_1$ is constructed by, for each block in $G_0$, greedily extending all segments it contains by the same length in each direction.  In order to avoid expanding tiny gaps, we greedily extend blocks in reverse order based on the number of sequences they contain. 

\subsubsection*{Filling in Missing Blocks}

We can only extend each block by the minimum length allowed for any segment it contains, and $G_1$ will therefore still not necessarily contain every position of the input sequences.  We complete the procedure by repeating the following for each block of $G_1$ until no bases are uncovered.  For each segment that can be individually extended in the right direction using the rules described above, we create a new block containing the bases that would have been added by extending these segments.  The same is repeated for the segments that can be extended to the left.  The (up to two) new blocks created in this step are then extended using the greedy extension rules.  Finally, all blocks in $G_2$ are merged together whenever possible to form the final output.  

\begin{algorithm}   
  \caption{HAL Interpolate($G, K$)}
  \label{alg:interpolate}
  \begin{algorithmic} 
    \STATE $L_{tot} \leftarrow $ sum of block lengths in $G$;
    \STATE $L_{block} \leftarrow L_{tot} / K$  
    \STATE $G' \leftarrow$ empty HAL graph
    \FOR {$s \in S$ ($S$ is the set of sequences in $G$)}
    \STATE $i \leftarrow 0$
    \WHILE{$i < \text{len}(s)$}
    \STATE $c \leftarrow $ block created from alignment column in $G$ containing $s[i]$
    \STATE $d \leftarrow$ max. distance between any base in $c$ and any base in $G'$
    \IF{$d \leq L_{block}$}
    \STATE $G' \leftarrow G' \cup c$
    \ENDIF
    \STATE $i \leftarrow i+ L_{block}$
    \ENDWHILE
    \ENDFOR
    \FOR {block $b \in G'$ (from largest to smallest)}
    \STATE maximally extend $b$ in both directions
    \ENDFOR
    \WHILE {$\exists$ position $x \in G | x \notin G'$}
    \STATE greedily create new block $b'$ from $x$
    \STATE $G' \leftarrow G' \cup b$
    \STATE maximally extend $b'$ in both directions
    \STATE greedily create blocks from bases in $G$ that are not in $G'$ and insert them into $G'$
    \ENDWHILE
    \FOR {all pairs of blocks $b_1, b_2 \in G'$}
    \IF{all $b_1$ and $b_2$ can be merged to form a valid gapless alignment}
    \STATE merge $b_1$ and $b_2$ in $G'$
    \ENDIF
    \ENDFOR
  \end{algorithmic}
\end{algorithm}

\subsubsection*{LOD Creation}

A series of levels of details can be generated from a source graph $G$ using the procedure outlined above.  The user specifies the scaling factor between two levels of detail and the maximum number of blocks $B$  to process per query, and algorithm iteratively generates coarser levels of detail until one is reached such that the entire alignment can be displayed in $B$ blocks.  An API is provided such that browser queries are directed to the most detailed level of detail such that the expected number of blocks returned is less than $B$.  The running time is $O(N\log{N})$ where $N$ is the number of bases in $G$ (also valid if $N$ is the number of segments in $G'$, which can be much smaller).  

\subsubsection*{Experimental Results}
\begin{figure} 
\centering
\includegraphics[width=400pt]{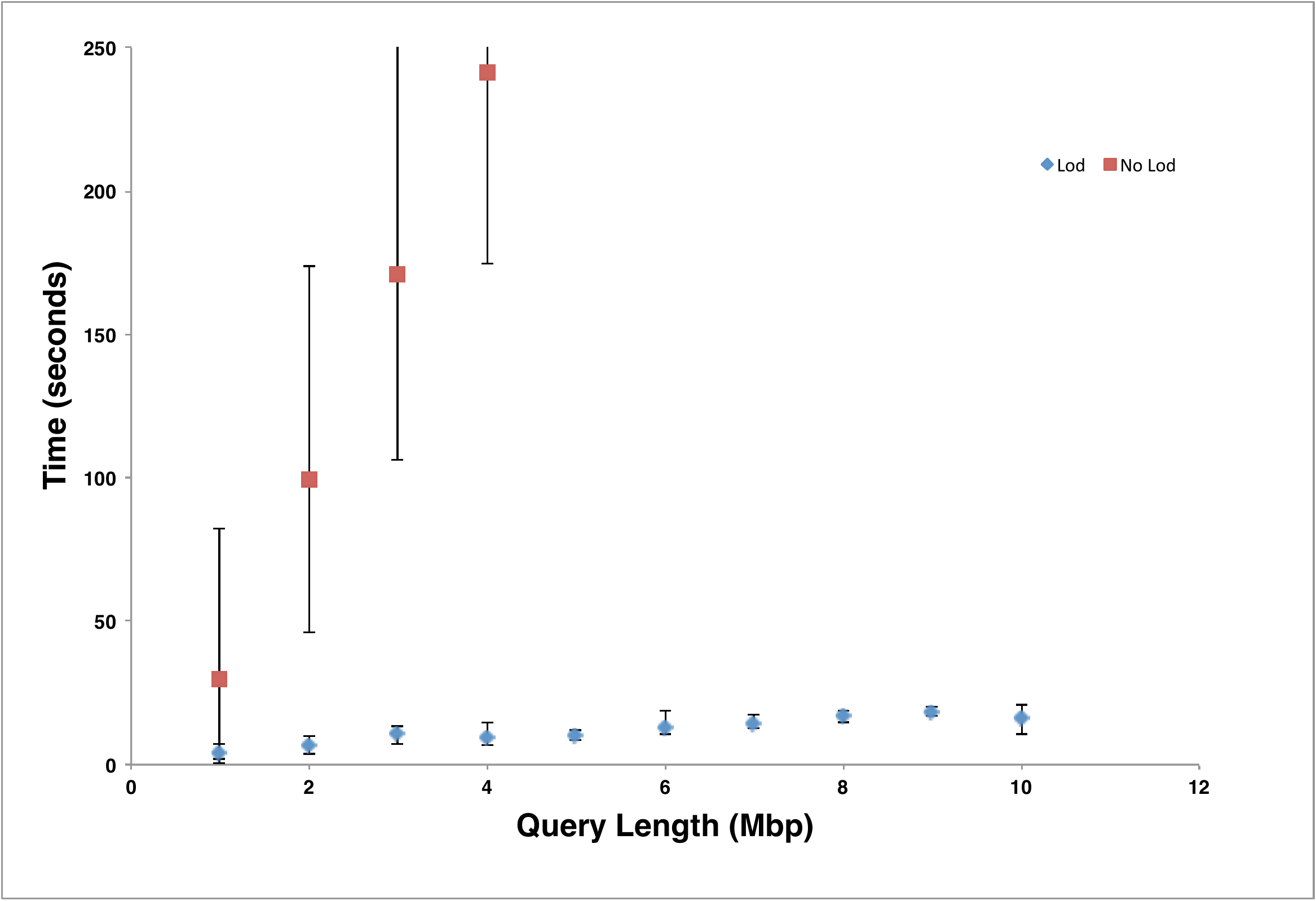}
\caption{(For online methods) Browser querying time with and without Level-Of-Details.}
\label{suppfig:lod}
\end{figure}

We assessed the practical impact of the level of detail generation by simulating 1000 random browser queries for alignments of four \textit{E. coli}/\textit{Shigella} strains (W 162011, KO11FL 162099, KO11FL 52593 and Ss53) to the \textit{E. coli} pangenome reference.  The size and location of the queries were uniformly distributed across the approximately 10 megabases of the reference genome.  The hub was hosted at the San Diego Supercomputer Center (La Jolla, California, USA), whereas the Browser server was located at the University of California Santa Cruz (Santa Cruz, California, USA), and no data was cached between individual queries.  Each random query was run the hub twice: once with levels of detail activated (Figure \ref{suppfig:lod}, blue dots) and once without (red dots).  The results are grouped, by query length, into bins of size 1000000 along the x-axis where bin N contains queries in the range (N-1000000, N].  The average time required for a query in each bin in seconds is reported in the figure, with the minimum and maximum query times shown in the error bars.  It is apparent from the chart that without levels of detail, queries quickly become impractical as the length approaches a megabase, but with levels of detail the time remains relatively constant for queries of any size.

\end{document}